\documentclass[pre,amsmath,amssymb]{revtex4}
\usepackage{hyperref}
\usepackage{colordvi}
\usepackage{epsfig}
\usepackage{bm}
\usepackage{graphicx}
\newcommand{\bbbZ}	{{\mathbb{Z}}} 			

\newcommand{\nn}{\nonumber} 
\newcommand{\be}{\begin{equation}} 
\newcommand{\ee}{\end{equation}}  
\newcommand{\bea}{\begin{eqnarray}}
\newcommand{\eea}{\end{eqnarray}}
\newcommand{\avg}[1]{\left< #1 \right>} 

\newcommand{\barr}      {\begin{array}}
\newcommand{\earr}      {\end{array}}

\begin{document}
\title{\bf Optimal trading strategies --- a time series approach}
\author{Peter A. Bebbington$^\dagger$ $^\ddagger$}
\affiliation{$^\dagger$Department of Physics and Astronomy, University College
London, Gower Street, London WC1E 6BT, U.K.}
\affiliation{$^\ddagger$ Centre for Doctoral Training in Financial Computing \&
Analytics, University College London, Malet Place, London WC1E 7JG, U.K.}
\author{Reimer K\"uhn*}
\affiliation{*Department of Mathematics, King's College London, Strand, London
WC2R 2LS, U.K.}

\date{\today}

\begin{abstract}
\hspace{6.0cm}\textbf{Abstract} \\
Motivated by recent advances in the spectral theory of auto-covariance matrices,
we are led to revisit a reformulation of Markowitz' mean-variance portfolio 
optimization approach in the time domain. In its simplest incarnation it applies to
a single traded asset and allows to find an optimal trading strategy which --- for 
a given return --- is minimally exposed to market price fluctuations. The model is 
initially investigated for a range of synthetic price processes, taken to be either 
second order stationary, or to exhibit second order stationary increments. Attention
is paid to consequences of estimating auto-covariance matrices from small finite 
samples, and auto-covariance matrix cleaning strategies to mitigate against these 
are investigated. Finally we apply our framework to real world data.
\end{abstract}
\maketitle

\section{Introduction}
When seeking an optimal strategy for capital allocation one can adopt a dynamic
programming approach that requires solving a Hamilton-Jacobi-Bellman or Bellman
equation \cite{Aurell2004, Muthuraman2006, Muthuraman2008, Lynch2010, Basak2010,
Brown2011, NBGarleanu2013, DeMiguel2014} to find such a strategy. An alternative 
approach, typically applied to single period problems, is mean-variance optimization, 
which forms the basis of Markowitz' portfolio optimization theory \cite{Markowitz1952}. 
This approach has a rich history in economic research and industrial practice 
\cite{Connor2010, Elton2010, Back2010, Bouchaud2006, Meucci2005}. One of the main 
reasons for its popularity is clearly its conceptual simplicity, which helps in building an intuition about the nature of risk and its relation to an investment's 
return.\\

The last couple of decades have seen many physicists becoming interested in this 
very same question \cite{Laloux1999, Plerou1999, Pafka2003, Ledoit2004a, Papp2005, 
MPotters2005B, Golosnoy2007, Chen2009, Still2010, Ledoit2012, FCaccioli2013, Caccioli2014}. 
Key issues addressed in these studies concern the effects that sampling noise is 
likely to have on the measurement of correlations or covariances in large portfolios, 
the way in which such sampling noise is going to affect the solution of a subsequent 
mean-variance portfolio optimization problem, and the design of methods to mitigate 
against adverse effects of such sampling noise. 

The bedrock of most of these studies is the theory of random sample covariance 
matrices \cite{Wish28}. Their spectral theory was pioneered by Mar\v{c}enko and 
Pastur \cite{MarcPast67} in the 1960's. It has indeed been observed that --- 
apart from a number of {\em large\/} eigenvalues --- the 
bulk of the spectrum of sample-covariance matrices of asset returns in various markets 
is very close to the form predicted by Mar\v{c}enko and Pastur for sample covariance 
matrices of i.i.d. random data; see e.g. \cite{Laloux1999, Plerou1999}. This type of 
comparison between market data and a null-model defined by random data could then 
be used to devise theory-guided ways of distinguishing between information and noise 
in market data, and thereby to devise methods to clean covariance matrices of asset 
returns for the purpose of their subsequent use in portfolio optimization, with the 
effect of improving risk-return characteristics  \cite{Laloux1999, Pafka2003, Ledoit2004a, 
Papp2005, MPotters2005B, Golosnoy2007, Chen2009, Still2010, Ledoit2012, FCaccioli2013, 
Caccioli2014}.

The present study was triggered by the fact that the spectral theory of sample {\em 
auto-covariance\/} matrices --- the analogue of \cite{MarcPast67} in the time domain ---
has recently become available \cite{Kuehn2012}. This leads us to revisit the analogue 
of Markowitz mean-variance optimization in the time domain \cite{Li2000}, which
in its simplest incarnation allows to find an optimal trading strategy for a single
traded asset over a finite (discrete) time horizon. We investigate this setup for 
a range of synthetic processes, taken to be either second order stationary, or to 
exhibit second order stationary increments, and we systematically study the effects
of sampling noise on optimal strategies and on risk-return characteristics. Finally
we apply our framework to daily returns of the S\&P500 index, and we explore how
results obtained for spectra of sample auto-covariance matrices obtained in 
\cite{Kuehn2012} could then be used as a guide to clean sample auto-covariance 
matrices in a spirit analogous to that used for sample-covariance matrices in the 
context of portfolio optimization. 

We note at the outset that we regard this as an exploratory study, and that we ignore 
economic factors such as discounting and agents' asymmetric perceptions of gains and losses in
the present paper. We expect that the primary area of application of our techniques would
be in the high-frequency domain, as return auto-correlations will be most prominent at short
times. We note, however, that much of our analysis is about effects of sampling noise on 
optimal trading strategies, which is relevant at {\em all\/} time scales, and thus also 
for weakly correlated data.  

The remainder of this paper is organized as follows. In Sect. II we briefly describe
Markowitz' approach to portfolio optimization, and its translation into the time domain.
In Sect. III we provide results for synthetic processes, and numerically investigate the 
influence of sampling noise on optimal strategies and risk-return profiles. In Sect. IV
we look at optimal trading strategies for empirical data, using the S\&P500 index as
an example and we investigate the effect of auto-covariance matrix cleaning on risk-return
profiles, based on comparing auto-covariance spectra for the S\&P500 and expected 
spectra for a process with uncorrelated increments. Sect. V is devoted to 
a final overview, and an outlook on promising future research directions.

\section{Portfolio Optimisation}

\subsection{The Markowitz Set-Up}
In the simplest version of mean-variance portfolio optimization one considers
a set of $N$ tradable assets $i=1,\dots,N$. It is usually assumed that these do 
not include complex financial instruments such as derivatives, options and futures.
An investor can take positions on these assets. We will use $\pi_i$ to denote
the position on asset $i$, using the convention that $\pi_i>0$ represents a
long position (buying the asset), whereas $\pi_i<0$ represents a short position
(selling asset). With $r_i$ denoting the (random) return on the $i$-th asset, 
the return on the entire portfolio with positions $\boldsymbol{\pi}=(\pi_1,
\pi_2,\dots,\pi_N)'$ is given by
\be
R(\bm\pi) = \sum_{i=1}^N \pi_i r_i = \bm \pi' \bm r\ ,
\ee
where $\bm r =(r_1,r_2,\dots,r_N)'$ is used to denote the vector of random returns and the prime indicates a transpose.

The optimal portfolio according to Markowitz is the one that minimizes the {\em
variance\/} of the portfolio return,
\be
\text{Var}[R(\bm\pi)] = \sum_{i,j=1}^N \pi_i\pi_j \avg{(r_i-\mu_i)(r_j-\mu_j)} 
= \sum_{i,j=1}^N \pi_i\pi_j\Sigma_{ij}\ ,
\label{Var}
\ee
subject to the constraint of a given expected portfolio return $\mu_P$
\be
\mu_P \equiv \avg{R(\bm\pi)} = \sum_{i=1}^N \pi_i \avg{r_i} =  \sum_{i=1}^N \pi_i \mu_i\ .
\ee 
In (\ref{Var}), $\Sigma=(\Sigma_{ij})$ is the covariance matrix of asset returns.

To put a scale to the problem, one usually imposes the normalization constraint
\be
\bm \pi' \bm 1 \equiv \sum_{i=1}^N \pi_i = 1\ .
\ee
Here $\bm 1 =(1,1,\dots,1)'$ denotes the $N$ dimensional vector with all components 
equal to 1. The minimization problem is solved using the method of Lagrange multipliers 
to take the constraint of expected return and normalization into account, i.e. one
looks the stationary point of the Lagrangian
\be
\mathcal{L}=\frac{1}{2} \bm \pi' \Sigma \bm\pi-\lambda_1(\boldsymbol{\pi}'{\bm
1}-1)-\lambda_2(\boldsymbol{\pi}'\boldsymbol{\mu}-\mu_P)
\label{eq:LangOptiMulti}
\ee
w.r.t variations of the $\pi_i, \lambda_1$ and $\lambda_2$. Elementary linear algebra
then entails that the optimal portfolio $\bm\pi^*$ takes the form
\be
\bm\pi^* = \lambda_1\Sigma^{-1} \bm 1 + \lambda_2 \Sigma^{-1} \bm \mu\ ,
\ee
with actual values of the Lagrange parameters $\lambda_1$ and $\lambda_2$ determined by 
the constraints.

\subsection{Translation into the Time-Domain}

The Markowitz portfolio optimization problem allows a fairly straightforward translation 
into the time-domain. To formulate it, assume that $X=(X_t)_{t\in\bbbZ}$ is the price process
for a single traded asset. Let $\pi_t$ denote the trading position that an investor takes on 
this asset at time $t$. As in the above we shall use the convention that $\pi_t>0$ represents 
a long position (buying the asset), whereas $\pi_t<0$ represents a short position (selling the 
asset).

The return of a trading strategy $\bm \pi=(\pi_1,\pi_2,\dots,\pi_T)'$ over a finite time horizon
of $T$ time steps for a realization $\bm x =(x_1,x_2,\dots,x_T)'$ of the price process can
be written as
\be
R_T(\bm\pi|x_0)=\sum_{t=1}^T\pi_t (x_0-x_t)\ .
\ee
In terms of these conventions the expected return $\mu_S$ of a trading strategy (conditioned 
on the initial price $x_0$) is
\be
\mu_S = \avg{R(\bm\pi|x_0)} =\sum_{t=1}^T\pi_t (x_0-\mu_t) = x_0 - \bm\pi'\bm\mu\ ,
\ee
where we have restricted ourselves in the second step to normalized trading strategies satisfying
$\bm\pi\bm 1 =1'$, and where $\mu_t=\avg{x_t}$ denotes the expected price at time $t$. 

It is worth remarking at the outset that $X$ could alternatively (and perhaps even more appropriately 
in the present context) be thought of as the log-price process, in which case $R_T(\bm\pi|x_0)$ would
be the log-return of the strategy $\bm \pi$. For the sake of simplicity and definiteness we shall stick
to the language of price processes and returns in what follows.

An optimal trading strategy in the spirit of Markowitz would then be a strategy which minimize 
the (conditional) variance
\be
\text{Var}[R_T(\bm\pi|x_0)] = \sum_{t,t'=1}^T \pi_t\pi_{t'} \avg{(x_t-\mu_t)(x_{t'}-\mu_{t'})}
= \sum_{t,t'=1}^T \pi_t\pi_{t'}\Sigma_{tt'}\ ,
\label{tVar}
\ee
subject to the constraints of normalization $\bm\pi'\bm 1 =1$ and given mean return $\bm\pi'\bm\mu
= x_0- \mu_S$. In (\ref{tVar}), the matrix $\Sigma = (\Sigma_{tt'})$ now denotes 
the {\em auto}-covariance matrix of the price process.

The algebraic side of the problem of finding an optimal trading strategy is now formally fully 
equivalent to that of finding an optimal portfolio, and the optimal strategy $\bm\pi^*$ takes the 
form
\be
\bm\pi^* = \lambda_1\Sigma^{-1} \bm 1 + \lambda_2 \Sigma^{-1} \bm \mu\ ,
\label{eq:OptT}
\ee
with $\Sigma$ now the {\em auto}-covariance matrix of the price process rather than the covariance 
matrix of portfolio returns. Actual values of the Lagrange parameters $\lambda_1$ and $\lambda_2$ are 
determined by the constraints as before.

It is well known, and indeed easily verified that the globally optimal solution which does not impose 
a restriction concerning the mean return is compactly given by
\be
\boldsymbol{\pi}^*_{\text{GO}}=\frac{\Sigma^{-1}{\bm 1}}{\bm 1' \Sigma^{-1} \bm 1}\ .
\label{eq:GMS}
\ee

The main problem facing both portfolio optimization \`a la Markowitz, and the mean-variance approach
to finding optimal trading strategies is that covariance matrices of portfolio returns or 
auto-covariance matrices of price processes of traded assets are not known, but need to be {\em estimated\/} 
from empirical market data. The effects of sampling noise in such estimation processes are well
studied in the case of portfolio optimization. As mentioned in the introduction, various strategies to mitigate 
against such effects --- typically guided by random matrix theory --- have been investigated in the past. 

By contrast, the corresponding random matrix theory for sample auto-covariance matrices that might be invoked 
for similar purposes for the problem of mean-variance formulations of optimal trading strategies has only 
recently become available \cite{Kuehn2012}. We shall address the issue of sampling noise in empirical data
and the use of spectral theory for the purpose of guiding the choice of ``cleaning"-strategies for 
auto-covariance matrices of market data below in Sect. IV. Before that we investigate the effects of sampling 
noise for some synthetic processes where comparison with known true auto-covariance matrices is possible.

\section{Results for Synthetic Price Processes}
In this section we evaluate the theory developed in the previous section for synthetic price processes. We 
begin by taking these processes to be either white noise processes or auto-regressive processes of order 1, 
and then move on to look at the situation where price{\em-increments} are modelled as white-noise and 
auto-regressive processes, respectively. For the white noise and auto-regressive price processes, the true
auto-covariance matrices are known, and analytical expressions for optimal trading strategies can be given.
We then look at the effects of sampling noise, using {\em estimates\/} of auto-covariance matrices for various 
values of the ratio of $\alpha=T/M$ of the length $T$ of the risk horizon (and thus the matrix dimension) and 
the sample size $M$ used to determine these estimates. The analytical expressions for the true auto-covariance 
matrices correspond to the $\alpha\to 0$-limit in these results.

\subsection{Synthetic Stationary Price Processes}
We first consider a price process with fluctuations around the trend $\delta x_t = x_t-\mu_t$ taken to be a 
Gaussian white noise process, i.e.  $\delta X_t\sim\mathcal{N}(0,\sigma^2)$. The true auto-covariance matrix 
in this case is proportional to the unit matrix, i.e. $\Sigma_{t,t'} = \sigma^2  \delta_{t,t'}$.

The globally optimal strategy (\ref{eq:GMS}) for a time horizon of length $T$ in this case  is then readily found 
to be
\begin{equation}
\pi_{t,\text{GO}}^*=\frac{\sigma^{-2}}{\sum_{t=1}^T{\sigma^{-2}}}= \frac{1}{T}\ .
\end{equation}
Thus, for a white noise process with variance $\sigma^2$ the optimal strategy $\boldsymbol{\pi}_\text{GO}^*=
(1/T,1/T,...,1/T)'$ is uniform over the time horizon $T$, and independent of the variance of the price process.
The analogous result for a Markowitz portfolio of uncorrelated assets is, of course, well known.

Let us next assume that price fluctuations around the trend are described by an AR(1) process, i.e. an auto-regressive 
process of order 1 of the form
\begin{equation}
\delta X_{t}=a\,\delta X_{t-1}+\Big (\sqrt{1-a^2}\Big )\xi_t\ ,
\label{eq:AR1Norm}
\end{equation}
in which $\xi_t \sim\mathcal{N}(0,1)$; for simplicity, we have normalized the process to exhibit fluctuations of
variance 1. The parameter $a$ in (\ref{eq:AR1Norm}) is required to satisfy $|a| < 1$ for fluctuations to be 
stationary. The auto-covariance function of this process is known to be given by
\be
\gamma(i)= \text{Cov}[\delta X_t \delta X_{t-i}] = a^{|i|}.
\ee
The auto-covariance matrix evaluated for a finite time horizon of length $T$ is thus a Toeplitz matrix of the form
\be
\boldsymbol{\Sigma}=\left (
\begin{array}{ccccccc}
1      & a        & a^2   &       &\cdots    & a^{T-1}\\
a      & 1        & a     &  a^2  &          & \vdots \\
a^2    & a        & 1     & a     &\ddots    &        \\   
       & a^2      & a     & 1     &\ddots    & a^2    \\
\vdots &          &\ddots &\ddots &\ddots    & a      \\
a^{T-1}&\cdots   &        & a^2   &  a       & 1      \\
\end{array}
\right )\ .
\label{eq:AutoCovTrue}
\ee
Its inverse is a tridiagonal matrix given by
\be
\boldsymbol{\Sigma^{-1}}=\frac{1}{1-a^2}\left (
\begin{array}{ccccccc}
1     & -a      &  0       & \cdots & \cdots  &   0   \\
-a    & 1+a^2   & -a       & \ddots &         & \vdots\\
0     & \ddots  & \ddots   & \ddots & \ddots  & \vdots\\   
\vdots& \ddots  & \ddots   & \ddots & \ddots  &  0    \\
\vdots&         & \ddots   & -a	    &  1+a^2  & -a    \\
0     & \cdots  & \cdots   &  0	    &   -a    & 1     \\
\end{array}
\right )\ .
\label{eq:IAutoCovTrue}
\ee
The globally optimal strategy (\ref{eq:GMS}) for a time horizon of length $T$ in this case  is then given by
\be
\bm\pi_\text{GO}^*= \lambda_1 (1,1-a,\dots,1-a,1)'\ ,
\ee
with $\lambda_1 = [2 + (T-2)(1-a)]^{-1}$ fixed by the normalization-constraint $\bm\pi'\bm 1 =1$. In this case
the globally optimal trading strategy turns out to be uniform apart from the two boundary terms. The white noise
result is clearly recovered as the $a\to 0$-limit of the present result for the AR(1) process as it should.

Solutions with constraints on the expected return can be given in closed form as well; they are simply obtained 
by inserting (\ref{eq:IAutoCovTrue}) into (\ref{eq:OptT}), with Lagrange parameters obtained by solving a pair of
linear constraint-equations; details will of course depend on assumptions concerning the drift, and we refrain
from writing them down explicitly. 

Fig. 1 shows optimal strategies for an AR(1) price process with parameter $a=0.8$, both for the global optimum as well 
as for cases with non-zero mean returns imposed. As can be seen from the figure, increasing the expected strategy 
return from $\mu_S=4.0\times 10^{-4}$ to $\mu_S=1.0\times 10^{-3}$ changes the optimal strategy (\ref{eq:OptT})
from one that is monotone decreasing over the risk-horizon to one which is monotone {\em increasing\/}, and starting 
in fact with a (short-)selling position at the initial time-step $t=1$.

\begin{figure}[ht!]
\includegraphics[scale=0.45]{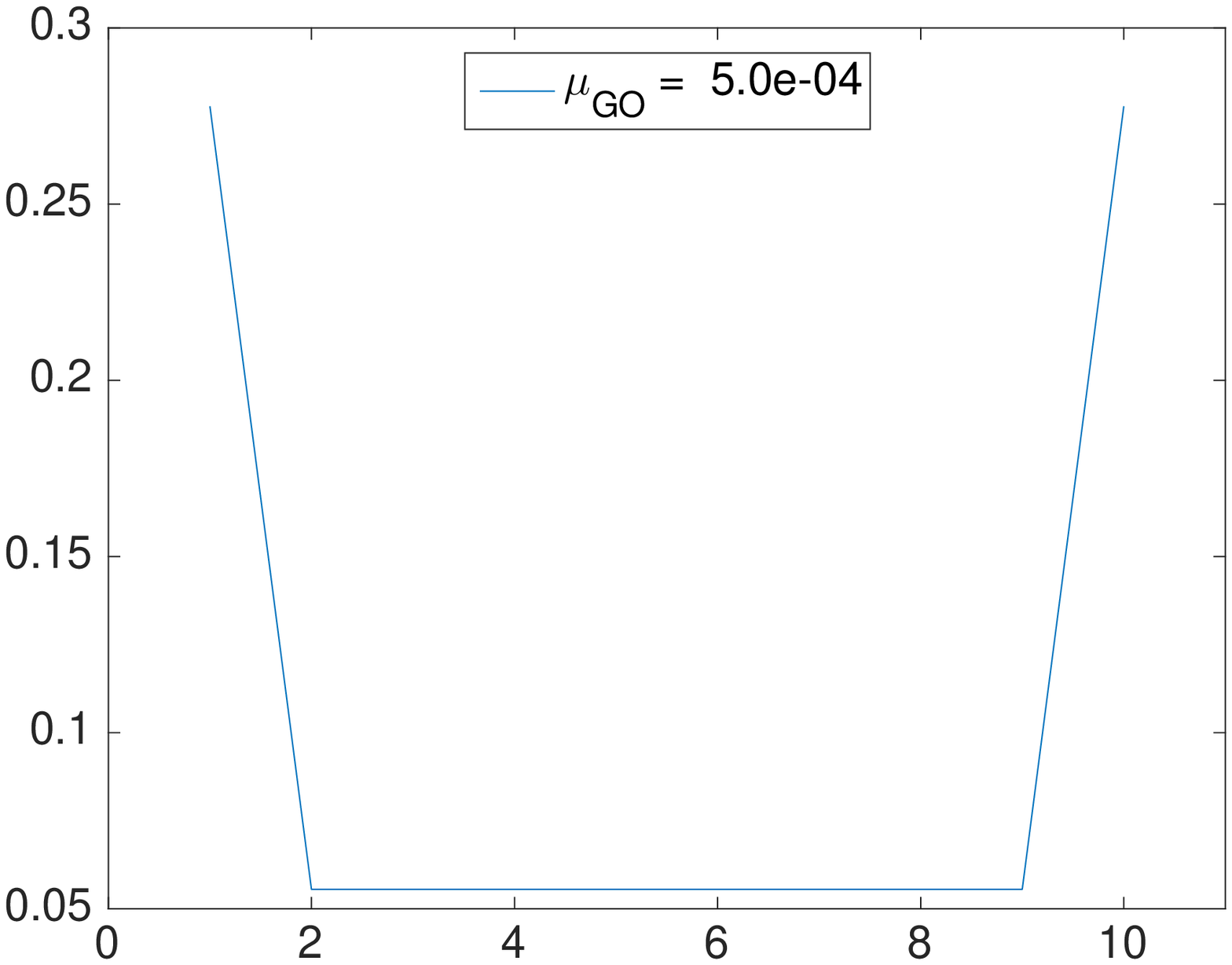}\hfill
\includegraphics[scale=0.435]{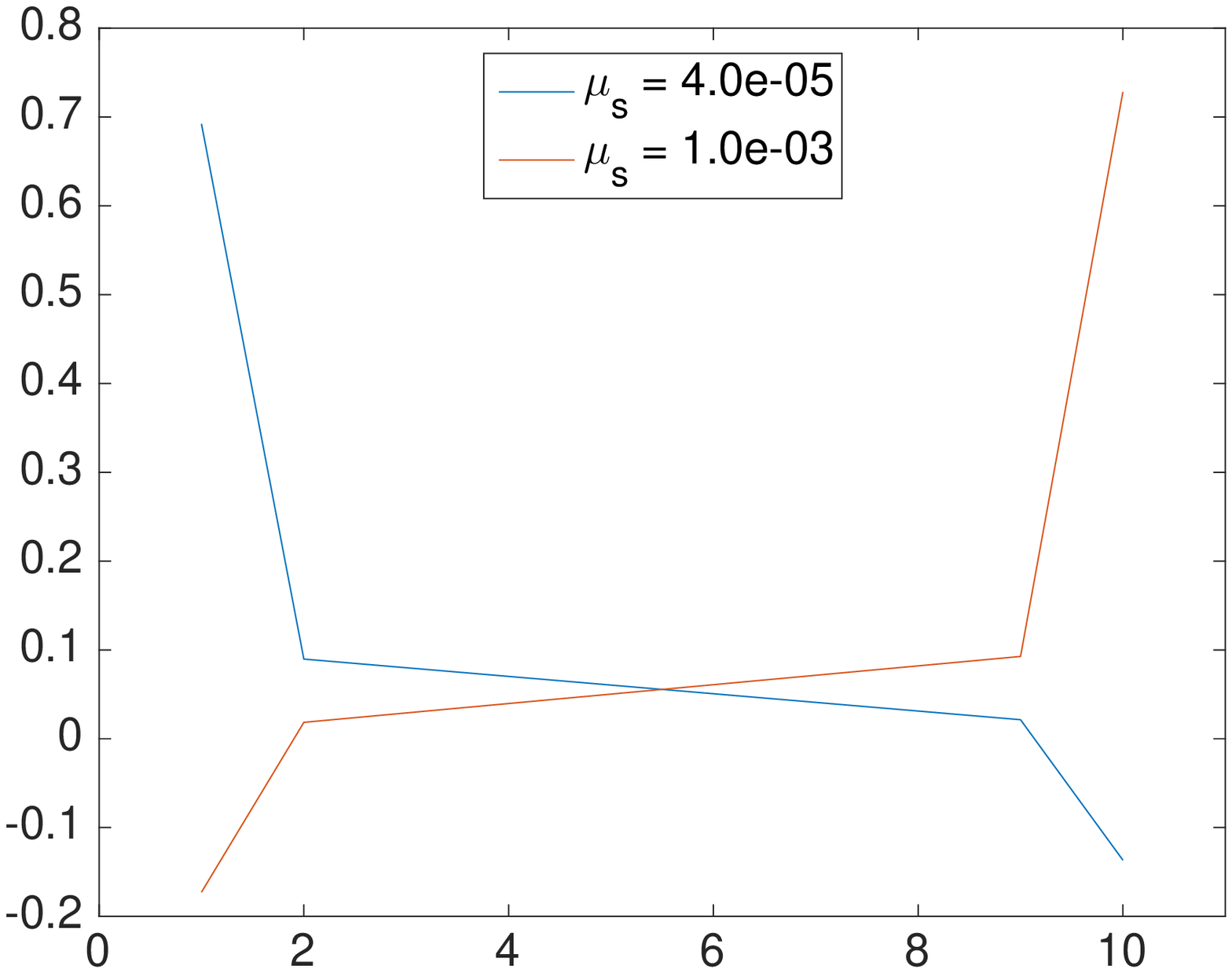}
\caption{Left panel: Globally optimal trading strategy for an AR(1) price process with $a=0.8$ over a risk 
horizon of $T=10$ time steps. Right panel: optimal strategies for a process with the same parameter
$a$ and a linear drift of the form $\mu_t = 10^{-4} t$, imposing expected strategy returns of
$\mu_S= 4 \times 10^{-5}$ (blue solid line) and $\mu_S=1 \times 10^{-3}$ (solid orange line).}
\end{figure}

\subsection{Synthetic Price Processes with Stationary Increments}
The stationarity assumption for the price process used in the previous subsection is clearly unrealistic, and
there is obviously need to go beyond that, if the methods discussed in the present investigation are to be useful in practice.

However, once the realm of stationarity is left, some structure is needed on a different level in order to make 
operational sense of estimating auto-covariance functions and the corresponding auto-covariance matrices defined 
over a finite time horizon. The structure we shall rely on here is based on the assumption that (fluctuations of) 
price-process can be described as having {\em stationary increments}. If one adopts the reading that the processes
considered here are actually log-price processes, the assumption of stationarity of their increments is actually 
a popular assumption in much of Mathematical Finance.

In what follows we assume that the (log-) price process $X=(X_t)$ exhibits stationary increments, i.e. that
\be
X_t=X_{t-1} + Y_t
\ee
with $Y_t= \avg{Y_t} + \delta Y_t = \mu_t-\mu_{t-1} + \delta Y_t$ with zero-mean fluctuations $\delta Y_t$. In 
terms of these conventions we can write the return of a strategy $\bm \pi =(\pi_t)$ for a given realization $\bm x$ 
as
\be
R_T(\bm\pi)=\sum_{t=1}^T\pi_t (x_0-x_t) =\sum_{t=1}^T\pi_t\Big[(\mu_0-\mu_t) - \sum_{\tau=1}^t \delta y_\tau\Big].
\ee
The expected return is given by the first contribution on the r.h.s, while the variance is
\be
\text{Var}[R_T(\bm\pi)] = \sum_{t,t'=1}^T \pi_t\pi_{t'} \Big[\sum_{\tau=1}^t \sum_{\tau'=1}^{t'} 
\avg{\delta y_\tau\delta y_{\tau'}}\Big].
\label{tVari}
\ee
This is of the same structure as (\ref{tVar}), with the auto-covariance matrix $\Sigma\equiv\Sigma^X=(\Sigma^X_{t,t'})$
of the non-stationary price process expressed in terms of the auto-covariance matrix $\Sigma^Y =(\Sigma^Y_{t,t'})$ of
the process of price increments as
\be
\Sigma^X_{t,t'} = \sum_{\tau=1}^t \sum_{\tau'=1}^{t'} \avg{\delta y_\tau\delta y_{\tau'}} = 
\sum_{\tau=1}^t \sum_{\tau'=1}^{t'} \Sigma^Y_{\tau,\tau'}\ .
\label{auto_statincr}
\ee
This relation between the auto-covariance matrices of process and the corresponding process of increments can be 
compactly expressed in matrix form as
\be
\Sigma^X = P \Sigma^Y P'\ , 
\label{PCP}
\ee
where $P$ is a lower triangular constant matrix of ones,  
\be
P = \left(\barr{ccccc}
    1 & 0 & 0 & \dots & 0 \\
    1 & 1 & 0 & \dots & 0 \\
    1 & 1 & 1 & \dots & 0 \\
 \vdots& \vdots &  \vdots & \ddots& \vdots\\
    1 & 1 & 1 & \dots & 1\earr\right )\ .
\ee

The mean variance approach to strategy optimization then yields optimal trading strategies of the form (\ref{eq:OptT}),
with the auto-covariance matrix $\Sigma =\Sigma^X$  of the price process expressed in terms of the auto-covariance 
matrix $\Sigma^Y$ of the process of stationary increments according to Eq. (\ref{PCP})

Taking the price increments to be a white noise process $\delta Y_t \sim\mathcal{N}(0,\sigma^2)$, we have
 $\Sigma^Y_{t,t'}= \sigma^2 \delta_{t,t'}$ so  $\Sigma^{-1}= \sigma^{-2} (PP')^{-1}$, where $(PP')^{-1}$ is found
to be of tridiagonal form,
\be
(PP')^{-1} = \left(\barr{cccccc}
    2 & -1 & 0 & 0 &\dots  & 0 \\
   -1 & 2  &-1 & 0 & \dots & 0 \\
   0 & -1 & 2  & -1 & \dots & 0 \\
   \vdots& & \ddots &\ddots &\ddots &\vdots\\
    0 & 0 & \dots & -1 &2 & -1\\
    0 & 0 & \dots & &-1& 1\earr\right )
\ee
The globally optimal strategy (\ref{eq:GMS}) in this case is then simply
\be
\bm\pi_\text{GO}^*= (1,0,0,\dots 0)'\ ,
\ee
i.e., it consists of taking a single long position at the initial time step.

If we assume an AR(1) process, of the form eq. (\ref{eq:AR1Norm}), for the fluctuations of the price increments, i.e.
\begin{equation}
\delta Y_{t}=a\,\delta Y_{t-1}+\Big (\sqrt{1-a^2}\Big )\xi_t\ ,
\end{equation}
then it is $\Sigma^Y$ which is given by Eq. (\ref{eq:AutoCovTrue}); it turns out that $\Sigma^{-1}= (P \Sigma^Y P')^{-1}$,
too, can be evaluated in closed form, giving
\be
\boldsymbol{\Sigma^{-1}}=\frac{1}{1-a^2}\left (
\begin{array}{ccccccccc}
  C     & -A^2     &  a         & 0         & \cdots     & \cdots &        & \cdots &  0   \\
-A^2    & 2B       &  -A^2      & a         &   0        &        &        &        &\vdots\\
 a      &   -A^2   & 2B         &  -A^2     & a          &  0     &	   &        &\vdots\\   
 0      & \ddots   & \ddots     & \ddots    & \ddots     & \ddots & \ddots &        &\vdots\\
\vdots  & \ddots   &            &           &            &        &        & \ddots &\vdots\\
\vdots  &          & \ddots     & \ddots    & \ddots     & \ddots & \ddots & \ddots &  0   \\
\vdots  &          &            &    0      &    a       &  -A^2  &   2B   &  -A^2  &  a   \\
\vdots  &          &            &           &   0   &    a        & -A^2   & C      & -A   \\
0       & \cdots   &\cdots      & \cdots    &   \cdots   &  0     &  a	   & -A     &   1  \\
\end{array}
\right )\ .\nn
\label{eq:IAutoCovTruei}
\ee
in which we use the abbreviations $A=1+a$, $B=1+aA$ and $C=1+A^2$.

In this case the globally optimal strategy (\ref{eq:GMS}) is of the form
\be
\bm\pi_\text{GO}^*= (1+a,-a,0,\dots,0)'\ ,
\ee
i.e. it consists of taking a single long position at the first time-step, which is then partially offset by a
short position at the second time step if $a>0$, whereas it is followed by a further long position if successive 
price increments are {\em anti}-correlated ($a<0$). Note that the solution for white noise increments is correctly 
recovered as the $a \to 0$-limit of the AR(1) results.

Once more, solutions with constraints on expected returns can be given in closed form; in analogy to the procedure
described for the case of stationary price processes, they are obtained by inserting (\ref{eq:IAutoCovTrue}) into 
(\ref{eq:OptT}), with Lagrange parameters obtained by solving a pair of linear constraint-equations. 

We find, and shall demonstrate below that the procedure predicts non-trivial changes of strategy as constraints on 
expected returns are varied. Once more, details will depend on assumptions concerning the drift, and we refrain from 
producing explicit equations here. We will report our analytical results alongside numerical results which take 
sampling errors arising from finite sample fluctuations on {\em estimated\/} auto-covariance matrices into account

\subsection{The Effects of Sampling Noise}
\label{sec:sampling-noise}

Having analytical results for synthetic price processes available allows one to estimate the effects of sampling
noise on optimal strategies and on risk return profiles. In practice, the analytic structure of an underlying price
process will not be known, and auto-covariance matrices will have to  be estimated on the basis of finite samples, i.e.
the design of optimal strategies will have to be based on {\em sample auto-covariance matrices\/} $\hat \Sigma$.

For a stationary price process, samples taken along a realization of the process can be taken to define the elements 
of $\hat\Sigma$ via 
\be
\hat \Sigma_{t,t'} = \frac{1}{M-1} \sum_{\mu=1}^M \delta x_{t+\mu} \delta x_{t'+\mu}\ .
\label{hSigma}
\ee 
This procedure introduces sampling noise; estimated auto-covariance matrix elements $\hat \Sigma_{t,t'}$ will exhibit
${\cal O} (M^{-1/2})$ fluctuations about their corresponding true counterparts $\Sigma_{t,t'}$. When assessing the effects
of sampling noise via the influence on spectra, one expects the relevant parameter to be the aspect ratio $\alpha = T/M$,
i.e. the ratio of the number of time-lags considered and the sample-size used to estimate matrix elements. We shall use 
this parameter in what follows to parametrize the influence of sampling noise, with the $\alpha\to 0$-limit corresponding
to the situation without sampling noise, i.e. with true asymptotic auto-covariances known.

\begin{figure}[ht!]
\includegraphics[scale=0.5]{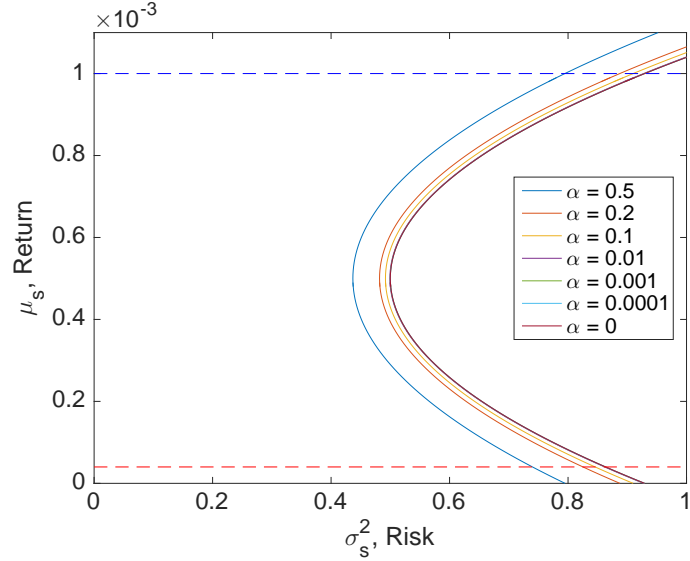}
\caption{Risk-return profile for an AR(1) price process with the same parameters as in Fig. \ref{fig:AR1_strat_mc1}
for various levels of sampling noise parameterized by $\alpha$. Results are obtained by averaging over $10 ^7$ samples 
as in Fig.\ref{fig:AR1_GM_strat_mc}. Note in particular that sampling noise leads to an {\em under-estimation of risk}.
The two horizontal dashed lines indicate two values of the target return for which optimal trading strategies are reported
in Fig \ref{fig:AR1_strat_mc1} below.}
\label{fig:RR_AR1_strat_mc}
\end{figure}

If the price process is not stationary, but has stationary increments, one can use Eqs. (\ref{auto_statincr}) and (\ref{PCP})
to express the auto-covariance matrix $\Sigma^X$ of the price process in terms of the auto-covariance matrix $\Sigma^Y$ of 
the process of price increments. For the latter it is legitimate to use an estimator by sampling along a realization, so
one can define $\hat \Sigma^X$  via
\be
\hat \Sigma^Y_{t,t'} = \frac{1}{M-1} \sum_{\mu=1}^M \delta y_{t+\mu} \delta y_{t'+\mu}
\label{hauto}
\ee 
and
\be
\hat\Sigma^X = P \hat\Sigma^Y P'\ . 
\label{hPCP}
\ee

In Fig. \ref{fig:RR_AR1_strat_mc} we show the risk-return profile for the case of an AR-1 price process for various
aspect ratios $\alpha$, ranging from $\alpha=0.5$ down to $\alpha = 10^{-4}$, with the noise-free case $\alpha=0$ also 
included. Note that sampling noise leads to a systematic underestimation of risk, though results quickly approach the 
noise-free limit as $\alpha$ becomes small. 

\begin{figure}[ht!]
\includegraphics[scale=0.5]{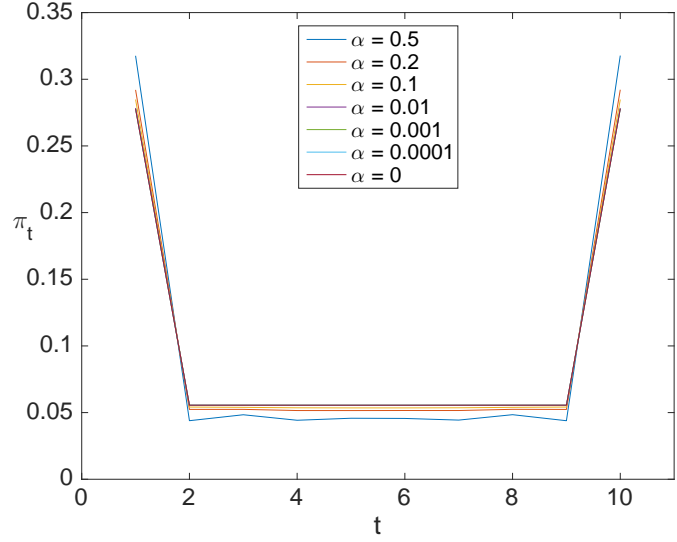}
\caption{Globally optimal trading strategies for an AR(1) price process with $a=0.8$ over a 
risk horizon of $T=10$ time steps, using {\em estimated\/} auto-covariance matrices. Data are
shown for various values of the ratio $\alpha=T/M$ of risk horizon and sample size $M$ used to
estimate auto-covariances according to Eq. (\ref{hSigma}): optimal strategies (with solid lines as guides
to the eye) are obtained by averaging over $10 ^7$ samples. Standard deviations are also shown; 
they rapidly decrease with $\alpha$. Results obtained for the {\em true} auto-covariance function 
(the $\alpha\to 0$-limit) are included for comparison. Note that average strategies obtained for 
finite samples are very close to the $\alpha=0$ results.}
\label{fig:AR1_GM_strat_mc}
\end{figure}

\begin{figure}[ht!]
\includegraphics[scale=0.45]{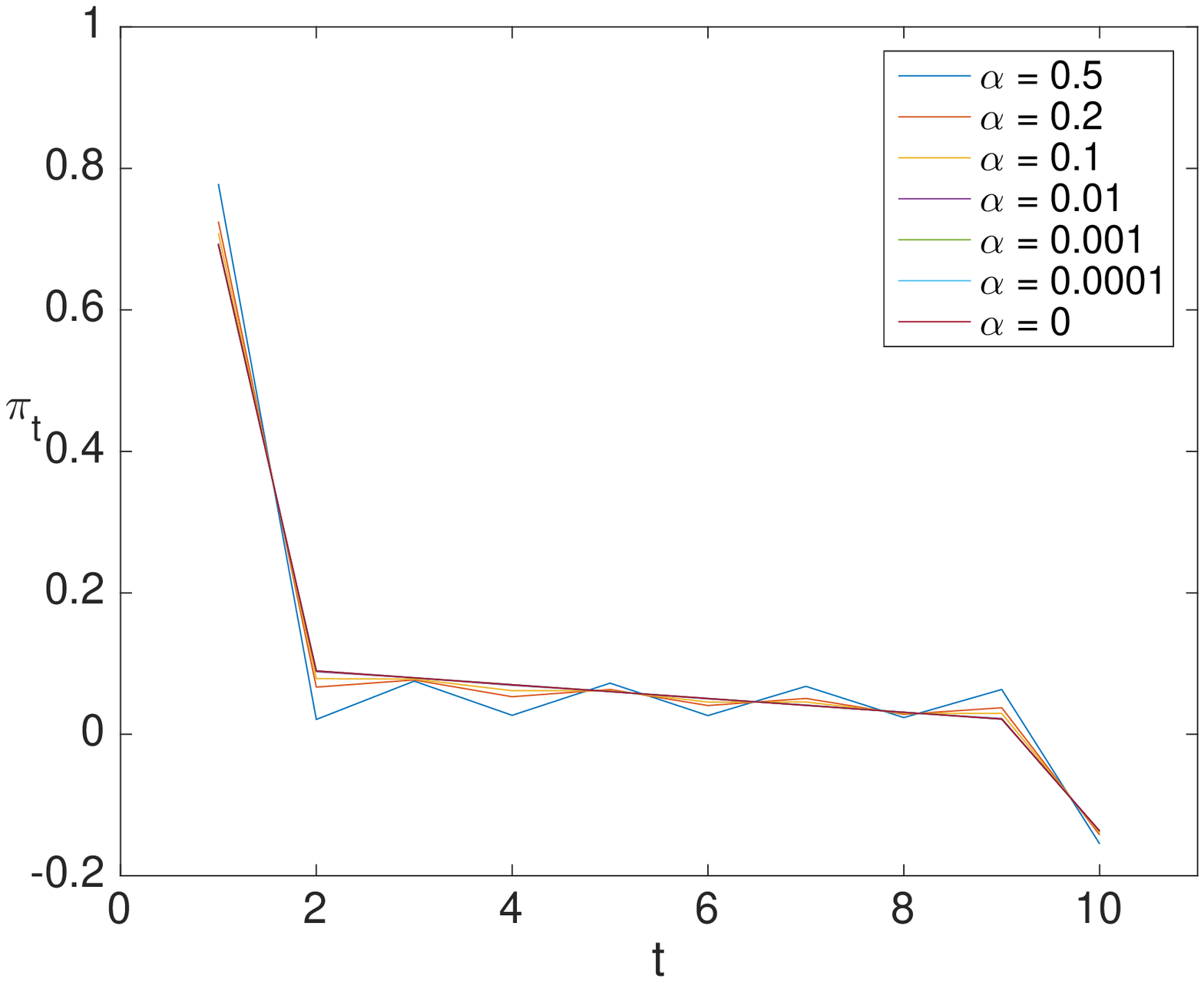}\hfill
\includegraphics[scale=0.45]{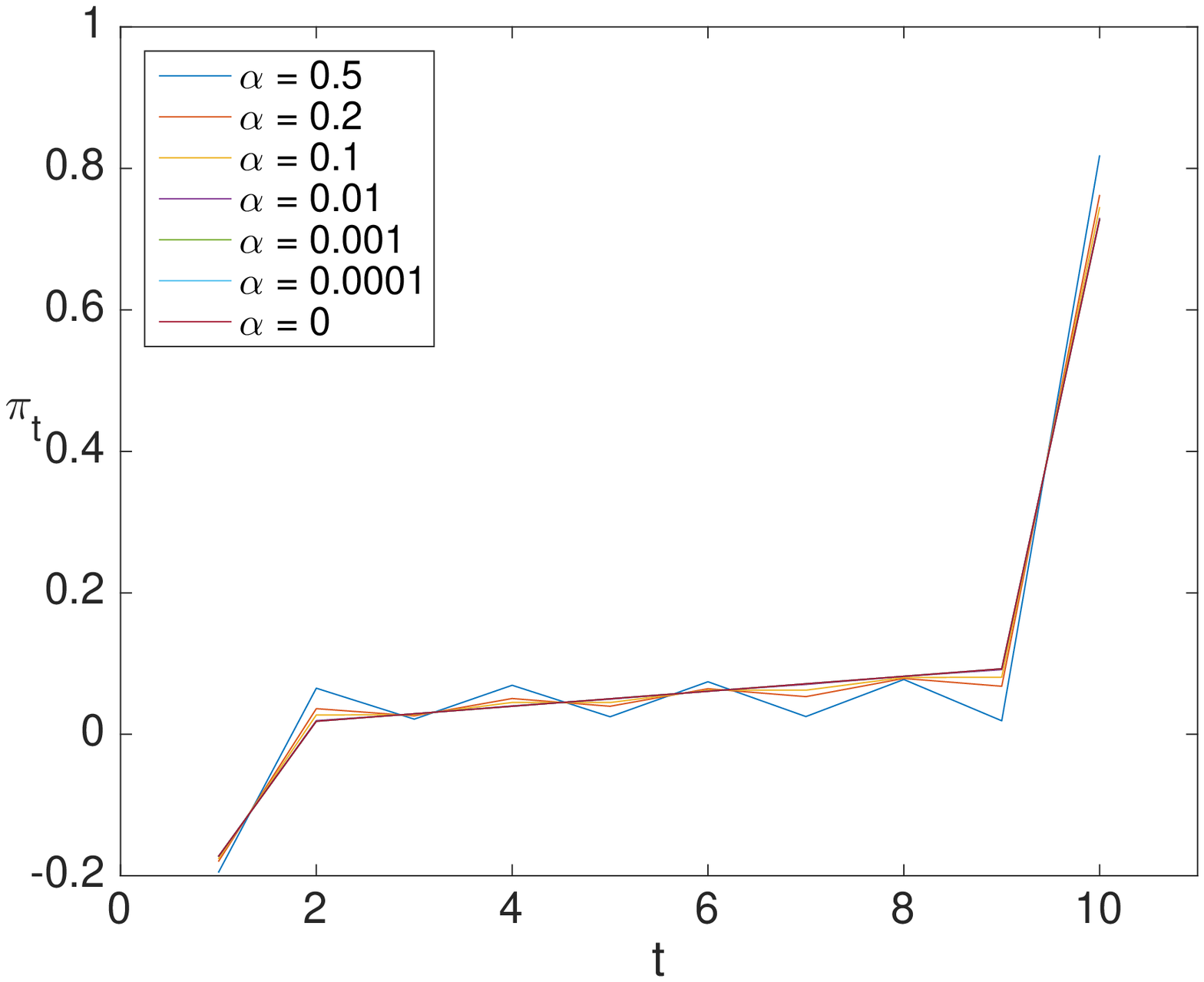}
\caption{Left panel: Optimal strategies for an AR(1)  process with $a=0.8$ and a linear drift of the form $\mu_t = 
10^{-4} t$ as in Fig. 1, with imposed expected strategy return of $\mu_S= 4 \times 10^{-5}$. Shown are average
trading strategies  for various levels of sampling noise parameterized by non-zero $\alpha$, obtained by averaging 
over $10^7$ samples. Average results are close to those obtained using true asymptotic auto-covariance matrices 
in the $\alpha\to 0$-limit, which are included for comparison. Right panel: optimal trading strategies for an AR(1) 
process with the same parameters as in the left panel, but now with $\mu_S= 1 \times 10^{-3}$.}
\label{fig:AR1_strat_mc1}
\end{figure}

Fig. \ref{fig:AR1_GM_strat_mc} exhibits the weights of the globally optimal (minimum risk) trading strategy for 
this process, while Fig. \ref{fig:AR1_strat_mc1} gives weights of optimal trading strategies for two different values 
of the target return (indicated by the two horizontal dashed lines in Fig. \ref{fig:RR_AR1_strat_mc}. In this case we 
assume a small drift $\mu_t = 10^{-4} t$ of the underlying price process. It is noticeable that an increase 
in the required target return leads to a qualitative change of the optimal strategy, with the larger target return 
requiring to take an initial short position at the beginning of the trading period. 

\begin{figure}[ht!]
\includegraphics[scale=0.45]{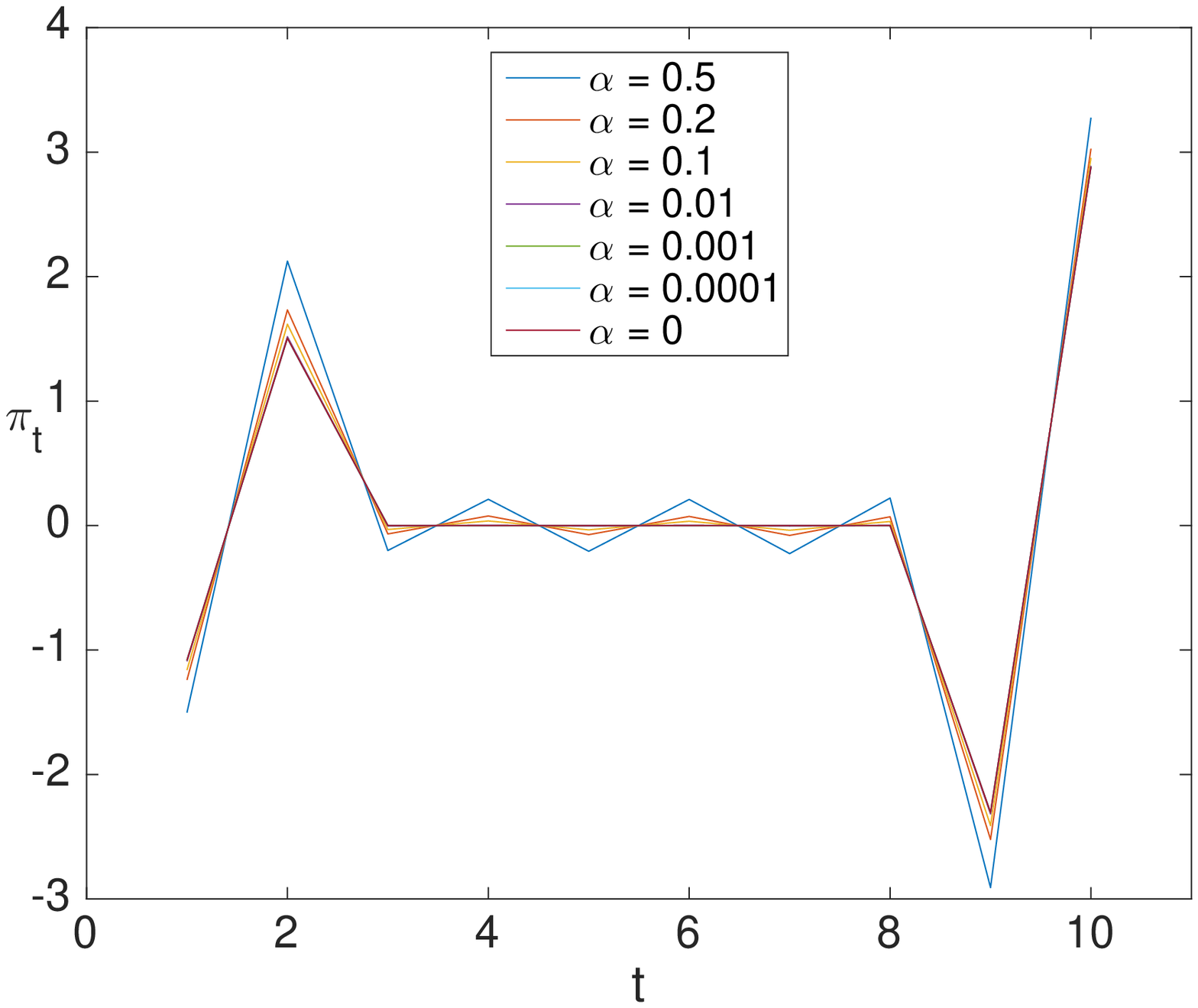}\hfil
\includegraphics[scale=0.45]{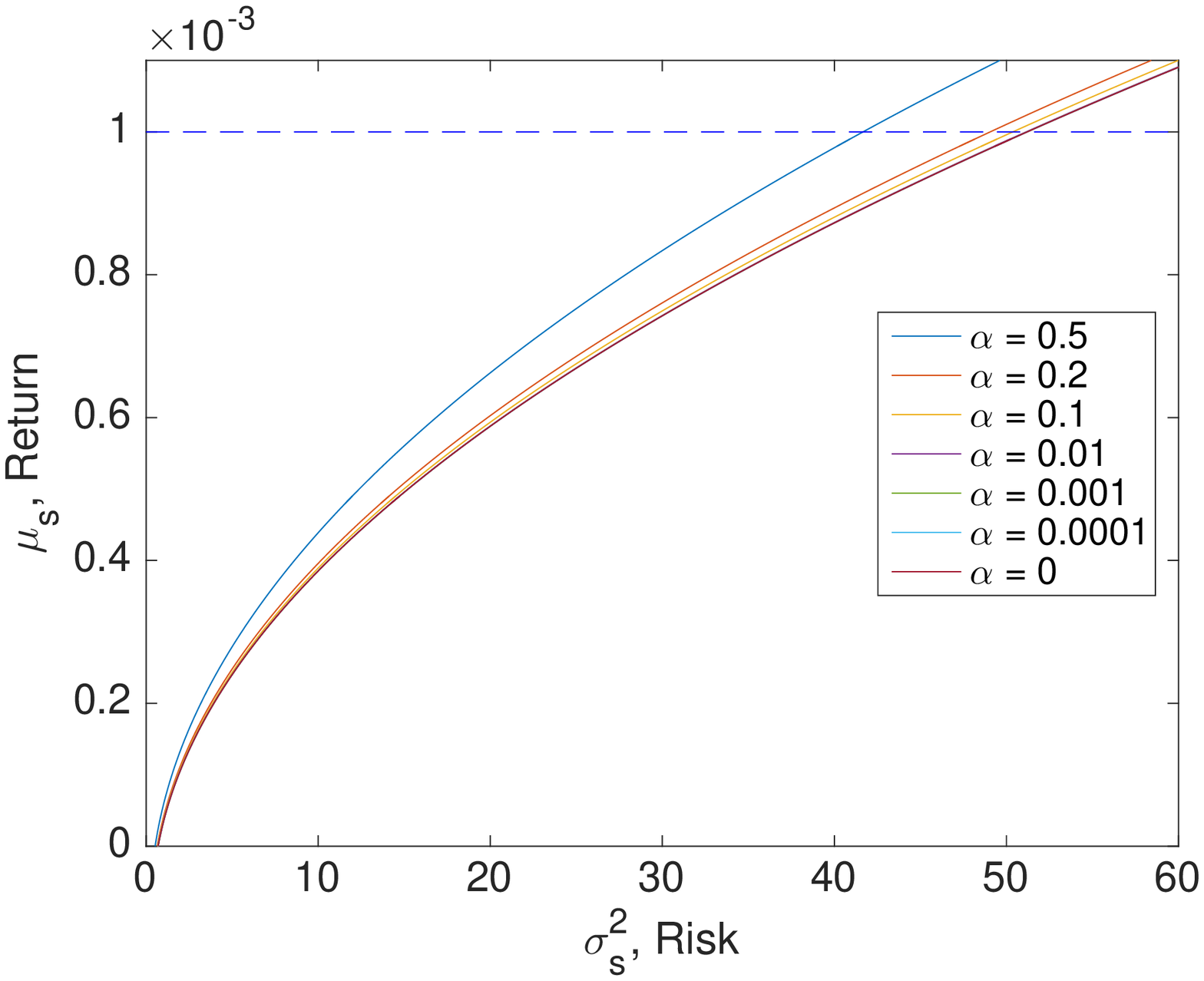}
\caption{Left: Optimal strategies for a setup where the fluctuations of the {\em price-increments} are described by an 
AR(1) process with $a=0.8$; a linear drift of the form $\mu_t = 10^{-4} t$ is assumed for the price process, and 
an expected strategy return of $\mu_S= 1 \times 10^{-3}$ is imposed. Shown are average trading strategies (solid lines) 
obtained by averaging over $10 ^7$ samples for various levels of sampling noise parameterized by non-zero $\alpha$. 
Average results are close to those obtained using true asymptotic auto-covariance matrices in the $\alpha\to 0$-limit,
which are included for comparison. Right: risk-return profile for this setup, with the horizontal dashed line indicating
the expected strategy return imposed in the data of the left panel. The right panel should be compared with Fig 
\ref{fig:RR_AR1_strat_mc}, which exhibits the risk return profile for an AR-1 price process.}
\label{fig:pooled_wav_x_7}
\end{figure}

Turning to the situation where we use an auto-regressive process to describe the statistics of price {\em increments}, 
we see from a comparison of Figs. \ref{fig:pooled_wav_x_7}  and \ref{fig:RR_AR1_strat_mc} that risk levels are 
significantly larger compared to the situation where the same underlying process describes the fluctuations of 
the price process itself.

This concludes our collection of results for synthetic price processes, where the underlying true auto-covariances are
known. We now turn to applying the framework to empirical data, where this is not the case.

\section{Empirical Data}

In what follows we apply our framework to empirical data, using daily adjusted close data of the S\&P500, spanning 
the period 03 Jan 1950 to 20 Apr 2015.

This is perhaps the point to notice that we are not advocating that using the variance of trading strategy returns
constitutes the best way of capturing risk in real market data. Indeed, given that market returns are known to have fat-tailed distributions, variance can at best be regarded as a proxy for risk. Howevever, our primary goal here is not to explore a wider family of possible risk measures, but rather to define a reformulation of the popular mean-variance optimization strategy in the time domain, and to begin investigating its properties. 

\subsection{The Spectrum of the S\&P500 Auto-Correlation Matrix}
Before turning to the evaluation of optimal trading strategies and risk-return profiles we shall have a look
at the spectrum of the auto-covariance matrices of the data, taking time windows of $T=50$, and sample sizes 
of $M=100$, hence $\alpha=0.5$. Auto-covariance matrices of the price process are obtained as described in 
Sect. \ref{sec:sampling-noise}, by first evaluating auto-covariances of the return process, assuming stationarity 
across individual sample-windows. In order to obtain meaningful statistics across the entire data set, we 
transform the return series in each time window to exhibit unit-variance increments, and then obtain 
auto-covariances of the thus normalized price process using the transformation Eq. (\ref{hPCP}).

\begin{figure}[hb!]
\includegraphics[scale=0.65]{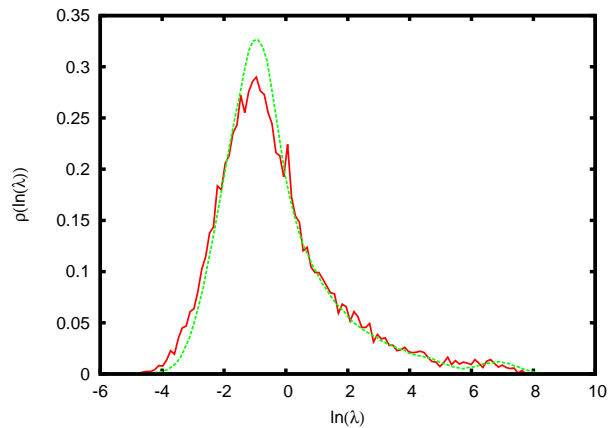}
\caption{Spectrum of the sample auto-covariance matrix of the S\&P500, normalized as described in the main 
text, using $T=50$ time lags and an aspect ratio $\alpha=0.5$, i.e. samples of size $M=100$ to define
the auto-covariances (red full line). Also shown is a comparison with the spectrum of an auto-covariance
matrix for a price process with {\em independent\/} unit variance increments (green dashed line). The two are 
remarkably close.}
\label{fig:SP500-WN}
\end{figure}

As can be seen in Fig. \ref{fig:SP500-WN}, where we plot the density of logarithms of eigenvalues,
the spectrum is {\em very broad\/}, spanning several orders of magnitude. For comparison we include the spectrum
for a process with {\em independent\/} unit variance increments using the same values of $T$ and $M$, and we notice 
that the two are remarkably close. This is not completely unanticipated, as it is one of the widely reported `stylized 
facts' in the field that return-series have very short correlation-times. We will use this type of spectral comparison 
below to inform the auto-covariance matrix cleaning strategy that we will use for the purpose of noise reduction.
 
\subsection{Optimal Trading Strategies and Auto-Covariance Matrix Cleaning}

\begin{figure}[h!]
\includegraphics[scale=0.45]{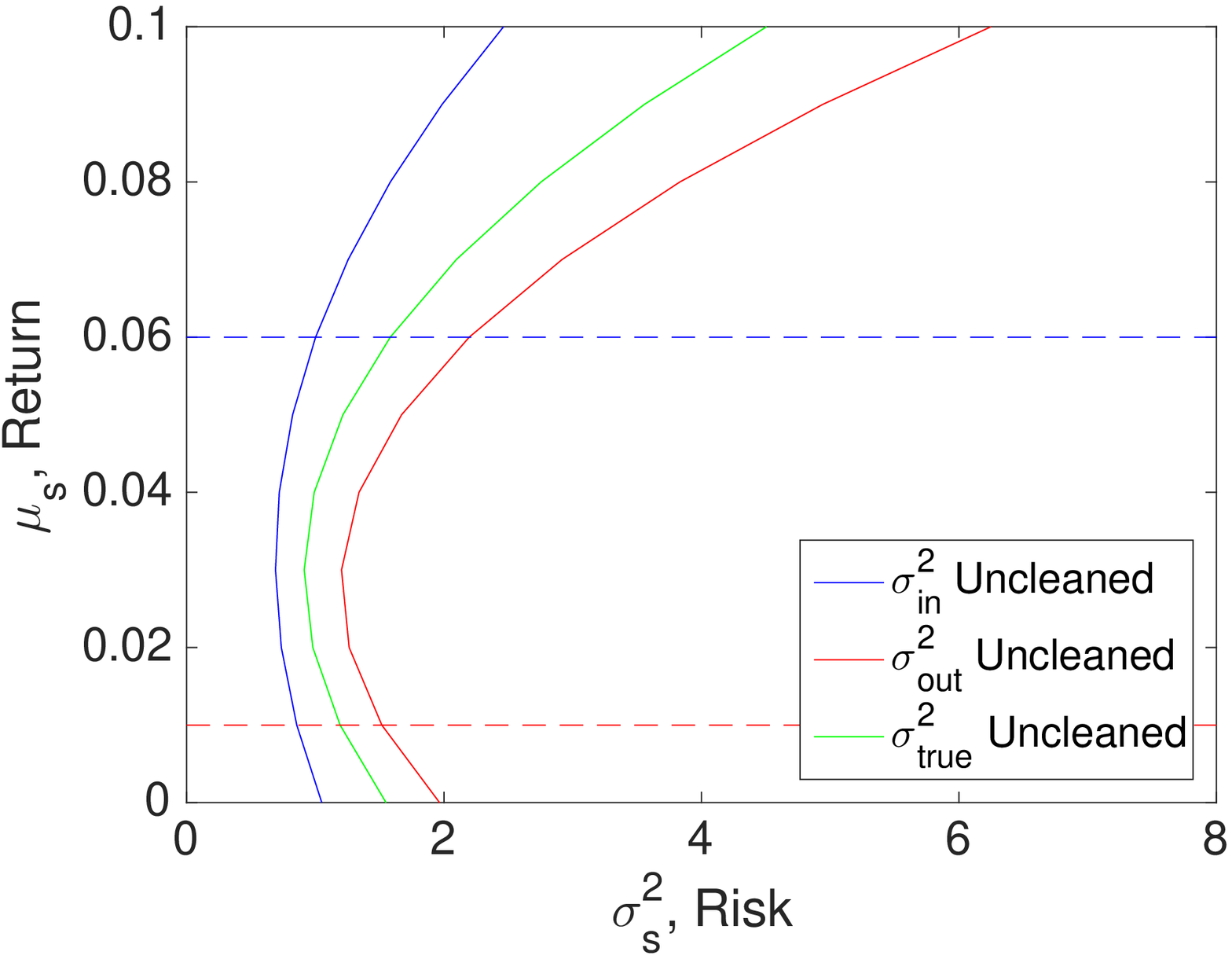}\hfill
\includegraphics[scale=0.45]{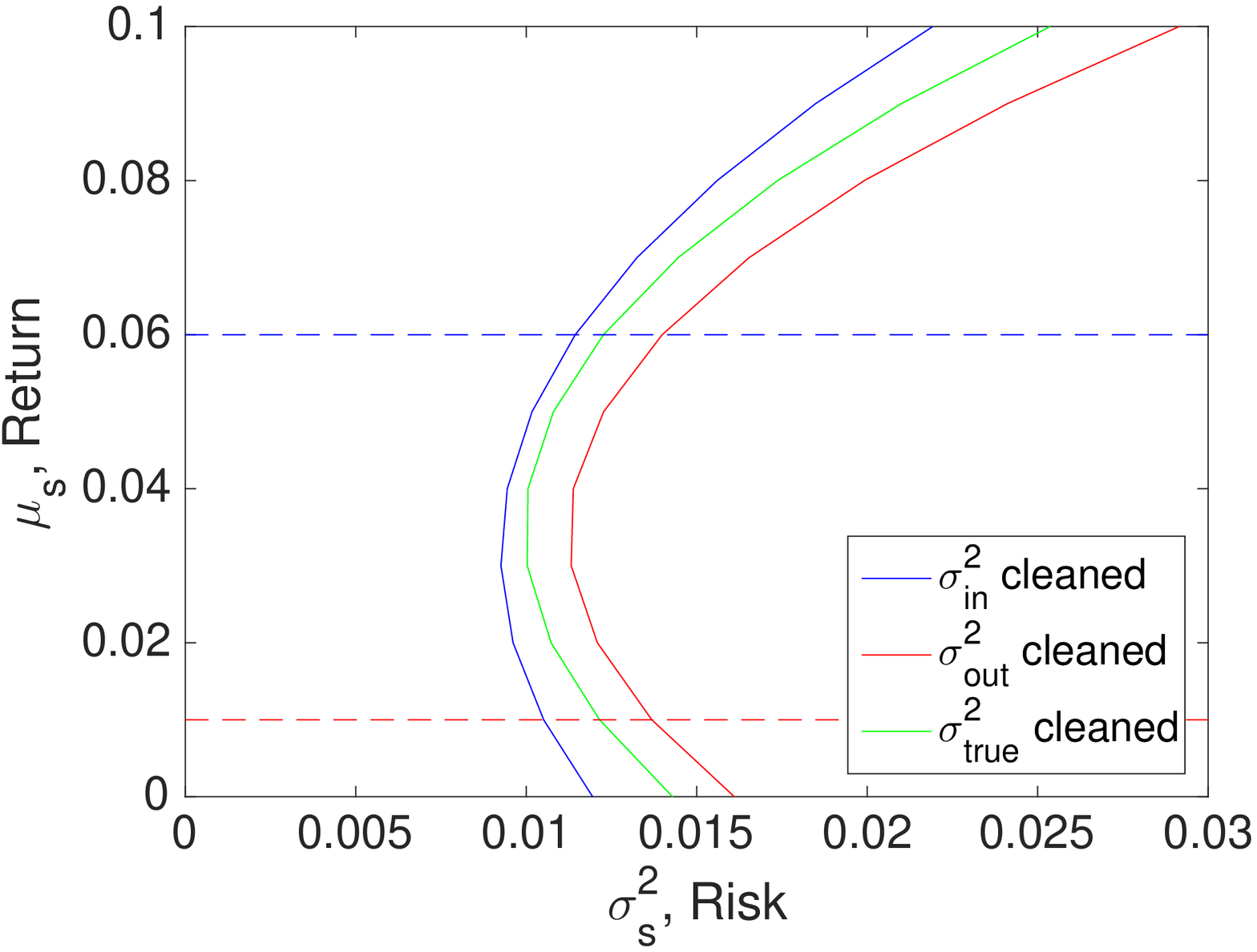}
\caption{Risk-return profile of optimal trading strategies on the S\&P500 data. Left: risk-return profile 
obtained from measured auto-covariance matrices. Right: risk-return profile obtained using cleaned versions
of auto-covariance matrices. Horizontal dashed lines denote target strategy returns $\mu_S$ for which optimal 
strategies are reported in Fig. \ref{fig:SP500-STR} below.}
\label{fig:SP500-RR}
\end{figure}

In Fig \ref{fig:SP500-RR} we report the risk-return characteristics for optimal trading strategies
on the S\&P500, using sample-auto-covariance matrices of $T=50$ time lags, and sample size $M=100$
as in Fig. \ref{fig:SP500-WN}. We report results obtained for auto-covariance matrices, as measured
via Eqs. (\ref{hauto}) and (\ref{hPCP}), and compare them with results obtained by applying
a cleaning strategy to these, which we shall describe below. We use  {\em realized\/} returns defined
by linear trends in each data window to compute risk-return profiles, and use conventions for in-sample 
risk, true risk and and out-of-sample risk as in \cite{Bouch2011}, taking the average 
auto-correlation matrix across the entire time series as a proxy for the true auto-correlation.
Note that the reduction of risk that can be obtained through cleaning is substantial. 

\begin{figure}[hb!]
\includegraphics[scale=0.45]{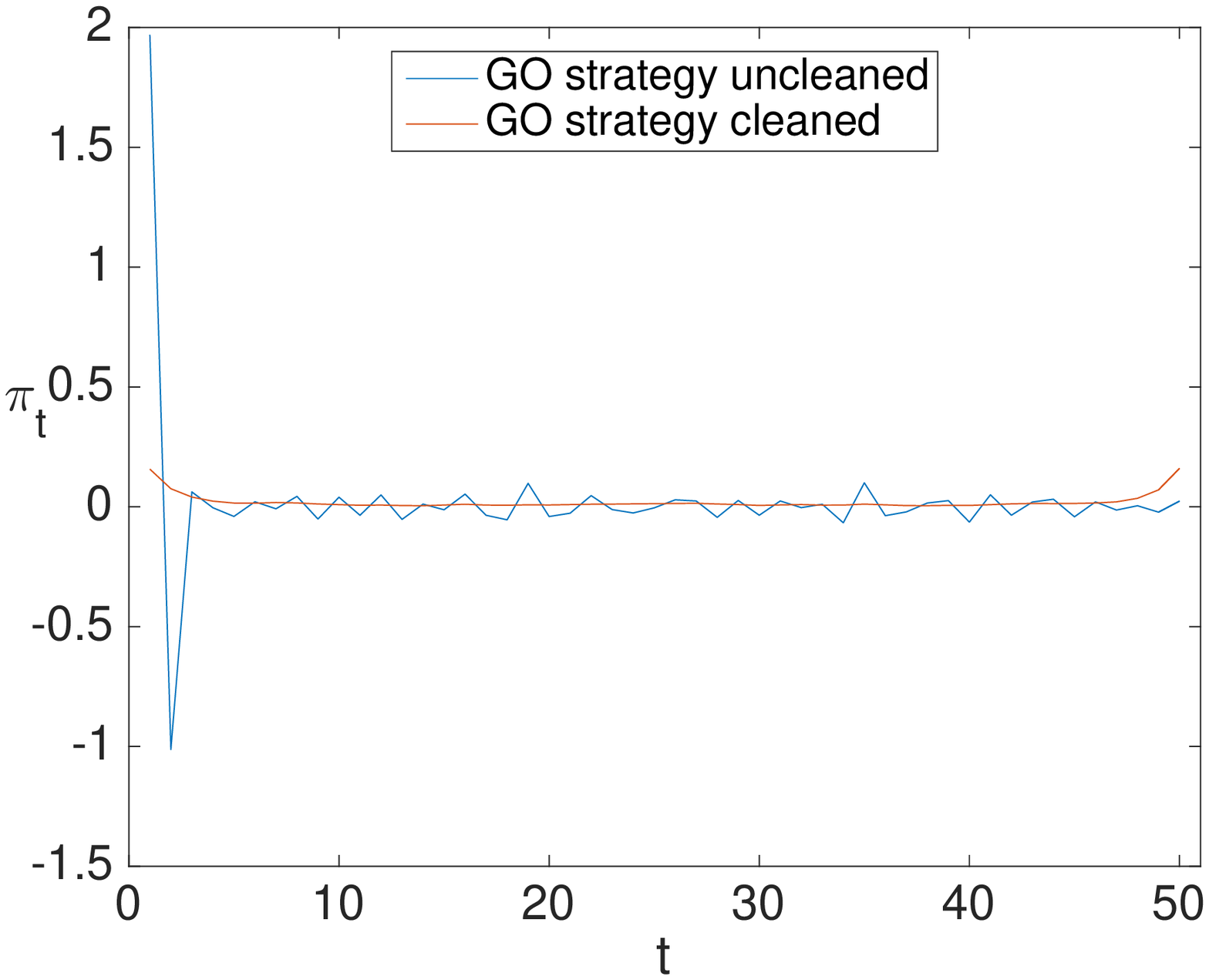}\hfill
\includegraphics[scale=0.45]{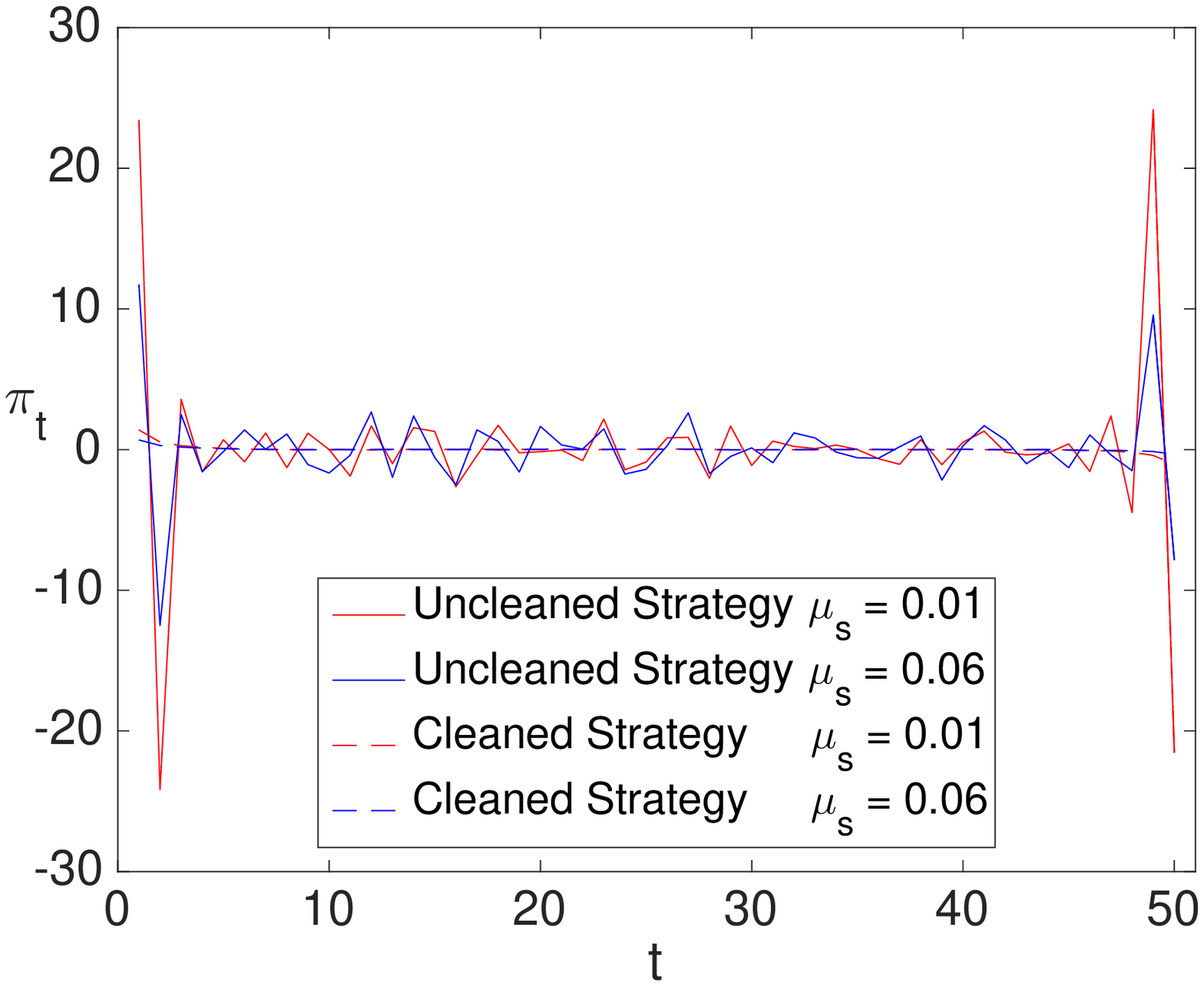}
\caption{Left: globally optimal strategy for the S\&P500, showing both results before and after cleaning. Right:
optimal strategies for the two target returns of $\mu_S=0.01$ and $0.06$ indicated in Fig. \ref{fig:SP500-RR} 
above.}
\label{fig:SP500-STR}
\end{figure}

Fig. \ref{fig:SP500-STR} exhibits optimal trading strategies for the S\&P500, showing both the minimal risk solution
and risk-optimal solutions for two different non-zero target strategy returns. Apart from the effect of reducing
risk, we find that the effect of cleaning is also to create strategies  that ``smoother" than those obtained without
cleaning.

Let us finally turn to the cleaning strategy that is used to obtain the data described above. In the context of
{\em covariance\/} matrices of financial data, strong similarities were observed between empirical correlation 
matrix spectra and the Mar\v{c}enko-Pastur law expected for high-dimensional uncorrelated data. One of the cleaning 
strategies that has been suggested due to such similarities is referred to as `clipping' \cite{Laloux1999, Bouch2011}. 
It analyses correlation matrices by performing a spectral decomposition, and regards the bulk of a sample correlation 
matrix spectrum, which resembles the Mar\v{c}enko-Pastur law, as noise. It then transforms correlation matrices by keeping 
large eigenvalues outside the bulk, and replacing those in the bulk by their average, thereby avoiding small eigenvalues in 
the transformed matrix.

In the present case, the phenomenology is rather different; there are no eigenvalues of the (normalized) sample 
auto-covariance matrices which can be regarded as lying significantly outside the bulk of the spectrum predicted for 
uncorrelated increments. So there would be no clear guidance coming from random matrix theory that could form the 
basis of a clipping-type procedure.

We therefore decided to apply a `shrinkage' procedure  to our data. To the best of our knowledge this procedure was first proposed by Stein \cite{Stei56}, and has recently found renewed interest in the Mathematical Statistics \cite{Ledoit2004a, Ledoit2012} and Econophysics \cite{Tum+07} communities. 

Based on the observation reported in Fig \ref{fig:SP500-WN} that the (normalized) auto-covariance spectra of the S\&P500 
and of a synthetic process with independent increments are indeed rather similar, we apply the shrinkage procedure
to the sample auto-covariance matrixes of the S\&P500 increments $\hat \Sigma^Y$, shrinking them towards a target
matrix $D$ given by the diagonal matrix of {\em variances} of the increments (which would indeed describe a process of 
independent increments), i.e. towards $D=\mbox{diag}(\{\hat\Sigma_{t,t}\})$, using the substitution rule
\be
\hat \Sigma^Y \leftarrow \delta D + (1-\delta) \hat \Sigma^Y\ ,
\ee
and transforming the shrunk $\hat \Sigma^Y$ thus obtained to define the cleaned estimate of $\hat \Sigma^X$ using 
the transformation Eq. (\ref{PCP}). The proper value for the parameter $\delta$ in this procedure is determined from
the data as described in \cite{Ledoit2004a, Ledoit2012}.

\section{Summary and Discussion}

To summarize, in the present paper we have a reformulation of Markowitz' mean-variance optimization in the 
time domain to obtain optimal trading strategies for a single traded asset over a finite discrete time horizon. Using
simple linear algebra, one obtains such optimal trading strategies as sequences of buy, hold, and sell instructions 
for that asset, which minimize the market fluctuations of the return generated by this sequence of instructions over 
a given time horizon, subject to suitable constraints. The procedure requires the auto-covariance matrix of the price
process (and estimates for expected prices) during the risk horizon as input.

We investigated this problem for a number of synthetic price processes, taken to be either second order stationary or be
described by second order stationary increments. Analytic expressions are given for the cases where the price and the 
return processes are described by i.i.d. or by auto-regressive fluctuations.

We compare analytic solutions with numerical results for situations where auto-covariance matrices have to be estimated
from finite samples, which is the situation typically encountered in practice. For the synthetic processes for which 
true auto-covariance matrices are known the effects of sampling noise on optimal strategies and on risk-return profiles
can thus be quantitatively assessed. We find that in general sampling noise leads to an underestimation of risk, but
that asymptotic results are well approximated when samples used to estimate auto-covariance matrices are sufficiently
large. A ratio $\alpha = T/M < 0.1$, i.e. sample sizes ten times the length of the risk-horizon appears to be desirable 
from this point of view.

From the financial point of view on the other hand, it is always desirable to use time series as short as possible for 
estimation, to avoid letting (possibly) outdated data influence current trading strategies. Small samples, however, 
increase the effects of sampling noise, and it is for this reason that cleaning strategies have an important role to 
play. Looking at the S\&P500 data, we found that (normalized) auto-covariance spectra closely resemble those one would 
expect for price processes with independent increments, and it is this observation that motivates our choice of target
matrix within a shrinkage cleaning strategy.

We observe that auto-covariance matrix cleaning gives rise to smoother trading strategies, and that it also leads to a 
reduction of risk in risk-return profiles.

A natural generalization of the present work would deal with a multi-period multi-asset version of a mean-variance 
formulation of optimal trading strategies. While some work has been done in this direction in the past (see. e.g. 
\cite{Li2000} and references therein) the solution presented in \cite{Li2000} remains somewhat formal, and restricted 
to the case without correlations in time. We are not aware of an investigation of the effects of sampling noise in
the multi-period multi-asset case. Indeed the spectral theory for that case which would be useful to motivate and design
cleaning strategies has not been developed as of now.

Another direction that could be pursued is to include higher moments of strategy-return distributions in measures of risk,
in order to better capture risk in the presence of fat-tailed return distributions. The translation into the time-domain,
as advocated in the present paper would in general involve $k$-point correlations of returns in time (where $k\geq 3$). 
Assessing sampling noise in such a situation would then clearly transcend the realm of random matrix theory
\begin{acknowledgments}
We would like to give warm thanks to the first authors PhD supervisors I.J. Ford and  F.M.C. Witte, the funding body EPSRC and the Centre for Doctoral Training in Financial Computing \& Analytics. 
\end{acknowledgments}
\bibliographystyle{unsrt}
\bibliography{refsn}
\end{document}